# Feasibility Evaluation of VANET using Directional-Location Aided Routing (D-LAR) Protocol

Ram Shringar Raw[1], Sanjoy Das[2], Nanhay Singh[1], Sanjeet Kumar[3], and Shailender Kumar[1]

[1]Ambedkar Institute of Advanced Communication Technologies & Research,
Delhi, India
rsrao08@yahoo.in, nsingh1973@gmail.com, shailenderkumar.verma@gmail.com

[2]School of Engineering and Technology, Galgotias University,
Uttar Pradesh, India
sdas.jnu@gmail.com

[3]Dr. Bhimrao Ambedkar University, Agra, India
sanjeet_hbti@rediffmail.com

**Abstract**
Vehicular Ad hoc Networks (VANETs) allow vehicles to form a self-organized network without any fixed infrastructure. VANETs have received wide attention and numerous research issues have been identified in the recent time. The design and implementation of efficient and scalable routing protocols for VANETs is a challenging task due to high dynamics and mobility constraints. In this paper, we have proposed D-LAR (Directional-Location Aided Routing), is an extension of Location Aided Routing (LAR) with Directional Routing (DIR) capability. D-LAR is a greedy approach based-position based routing protocol to forward packet to the node present in request zone within the transmission range of the source node as most suitable next-hop node. We have justified the feasibility of our proposed protocol for VANET.

## 1. Introduction

Vehicular Ad hoc Network (VANET) is a rapidly emerging new class of mobile ad hoc networks. VANET consists of a large number of vehicles providing connectivity to each other. We assume that vehicles move in every direction with high mobility. Neighboring vehicles that are within a transmission range directly communicate over a wireless links. If source and destination vehicle are not in direct communication range of each other, then they communicate through intermediate node in multi-hop fashion.

VANETs have a large potential to improve the traffic safety and travel comfort of drivers and passengers [1]. For example vehicles can communicate detour, traffic congestion, and accident information with the nearby vehicles to reduce the traffic jam near the affected areas. Rescue vehicles could instantly receive exact location information of the accident site to reach there faster. Toll could be automatically paid, traffic signals equipped with communication equipments could more accurately control intersection traffic. Although fixed infrastructure is not required for VANET, however fixed network nodes may be used in the form of roadside units. These roadside units open up a wide variety of services for vehicular ad hoc networks such as serving as a gateway to the Internet, serving up geographical data etc [2].

The work in this paper involves two steps. First, we explain the proposed routing protocol in VANET which is very essential to establish a path for packet transmission between nodes. In second step, we evaluate the feasibility of VANET for the dense traffic scenario using proposed routing protocol [3]. Our work only focuses on a densely connected city network.

Unlike traditional ad hoc and other routing protocols, position-based routing protocols present challenging and interesting properties of VANETs [4, 5]. A position-based routing protocol does not require any information on the global topology, but uses the local information of neighboring nodes that restricted to the transmission range of any forwarding node. Due to this restrictions, it gives low overhead of their creation and maintenance. Generally position-based routing is based on greedy forwarding scheme that guarantees loop-free operation. The local information about the physical location of nodes can be provided by the Global Positioning System (GPS), if vehicular nodes are equipped with a GPS receiver [6, 7]. Over the last few years, there have been numerous variations of position-based routing protocols such as LAR and DIR protocols examined in the literature [8, 9].

In this work, we study the performance of both LAR and DIR protocols and propose a new protocol based on greedy forwarding approach, that we call Directional-

Location Aided Routing (D-LAR) protocol. In this protocol, we first draw a straight line between source and destination. Then the packet is forwarded in the request zone to the direct neighbor having direction closest to the line drawn between source and destination.

The rest of the paper is organized as follows. We discuss the related work in section 2. In section 3, the design of D-LAR routing protocol is introduced. Section 4 presents the performance analysis of the proposed protocol. Finally, we conclude the paper in section 5.

## 2. Related Work

Ad-hoc routing protocols are classified into two categories: topology-based routing protocols and position-based routing protocols. Topology-based routing protocols are not more suitable in VANETs where the network topology is changing frequently. To cope with the dynamic network, position-based routing protocols are used. In this section, we introduce two existing position-based routing protocols like DIR and LAR protocols.

### 2.1 DIrectional Routing (DIR)

Kranakis [10] proposed the Directional Routing (DIR) (referred as the Compass Routing) is based on the greedy forwarding method in which the source node uses the position information of the destination node to calculate its direction. In DIR protocol next-hop neighbor node is decided through unicast forwarding by using the position information of the sender node, its next-hop neighbor nodes, and the destination node. Then the message is forwarded to the nearest neighbor having direction closest to the line drawn between source and destination. Thus a message is forwarded to the neighboring node minimizing the angle between itself, the previous node, and the destination.

The performance of DIR depends on the network density. DIR have high packet delivery rates for dense vehicular networks and low delivery rates for sparse vehicular networks. In the Fig. 1 (a), source node S has five neighbors in it transmission range. Among these nodes, node F has the closest direction (node F has minimum angle among all the neighbors within the communication range) to line *SD* and is the selected next-hop node for further packet transmission. Figure 1 (b), shows the direction of path selection by DIR protocol.

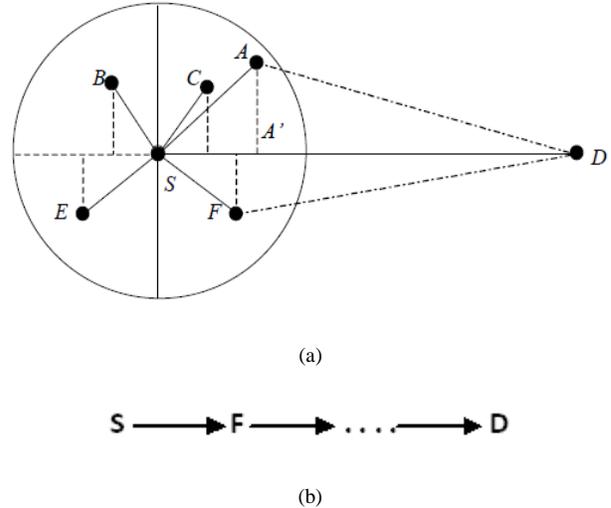

Fig. 1 (a) DIR forwarding method, (b) Path selected by DIR

### 2.2 Location-Aided Routing (LAR)

Several routing protocols have already been proposed for ad hoc networks. Young-BaeKo and Nitin H. Vaidya proposed the LAR protocol [8]. LAR is an on-demand routing protocol whose operation is similar to DSR (Dynamic Source Routing) [11]. As like DSR, LAR utilizes location information using GPS to limit the area for discovering a new route to a smaller "request zone". The location information is used to reduce routing overhead and improve the performance of routing protocols for ad hoc networks. In LAR, instead of flooding the route requests into the entire network, only those nodes in the request zone will forward packets.

In LAR, two zones are defined: expected zone and request zone [12]. The source node estimates a circular area (expected zone) in which the destination is expected to be found at the current time. The position and the size of the circle are calculated based on the location information and speed of destination ($v$) at time $t_0$. The request zone is the smallest rectangular region that includes the expected zone of radius R (= $v$ ($t_1$-$t_0$)) and the source nodes that should forward the route request packets to the destination.

The coordinates of the four corners of the request zone are included in the route request packet when initiating the route discovery process. RREQ broadcast is limited to this request zone. Thus, when the node in the request zone receives RREQ, it forwards the packet normally. However when a node which is not in the request zone receives an RREQ, it drops the packet. Therefore, the overhead of control packets is reduced. In Fig. 2, for example, if node *I* receives the RREQ from another node, node *I* forwards the RREQ to its direct neighbors because it is located in the request zone. However, when node *J* which is not in the request zone receives an RREQ, it discards the request.

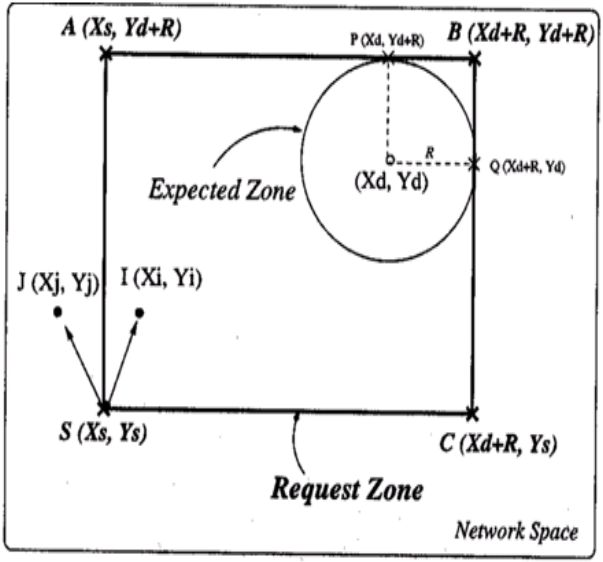

Fig. 2 Request and expected zones in LAR [12]

## 3. Proposed Scheme

This section is divided into two parts: Section 3.1 describes the assumptions of the proposed model while section 3.2 introduces our proposed scheme.

3.1 System Model

Since this scheme is improved based on DIR and LAR position-based routing protocol, some assumptions are made as follows:
- A large vehicular ad hoc network is considered.
- Vehicles forward packet only in ad hoc mode.
- Each vehicle is equipped with GPS to obtain the position information of itself.
- Every vehicle finds its own coordinates through receiver.
- The source vehicle forwards its own position through beacon message so that the neighboring vehicles forward packets to the destination.
- Impact of signals locked by obstacles or buildings not included.

3.2 Directional-Location Aided Routing (D-LAR) Protocol

In this section, we propose a Directional Location-Aided Routing (D-LAR) scheme to improve the performance of LAR scheme. D-LAR uses the advantage of DIR and LAR both. In DIR, the source node forward packets in a geographical area towards the direction of the destination node. LAR uses this directional forwarding method to forward the packet between source and destination node. This scheme is based on greedy forwarding approach which uses the location and direction of motion of their neighbors to select the next-hop node for further transmission.

The packet delivery ratio in position-based routing protocols depends on the position information of nodes obtained from the GPS. Each node finds its own location information once in every beacon interval. In high mobility network such as VANET, beacon interval has to be small to find more accurate position information but it gives high communication overhead that affect the overall performance of the network. Therefore, we need a method to find current position and direction of nodes towards destination in highly mobile network with low control packets overhead. So our proposed routing protocol finds the position of nodes using direction of nodes within beacon interval during packet transmission.

In D-LAR, the sender node selects the next-hop forwarding node having direction closest to the straight line drawn between source and destination. D-LAR selects the next-hop node from the nodes running in the same direction as the forwarding node because moving direction can help to build a stable route between forwarding and next-hop node. Therefore, a message is forwarded to the next-hop node in the request zone minimizing the angle between itself, the previous node and the destination node.

In D-LAR, the straight line drawn between source and destination node in the requested zone is used to determine the angle of the nodes within the transmission range. Then D-LAR select the node having minimum angle as a next-hop node for further transmission. In this method, angle (direction closest to the straight line, *SD*) of nodes can be determined through directional routing scheme as given below:

The positions of the two mobile nodes *S* and *A* in Fig. 3 are $(x_1, y_1)$ and $(x_2, y_2)$ respectively. In Fig. 3, when node *A* receives the RREQ from source node *S*, node *A* will calculate the distance *d* between the source node and itself as

$$d = \sqrt{(x_2 - x_1)^2 + (y_2 - y_1)^2} \qquad (1)$$

Similarly, the angle $\theta$ for node *A* with respect to straight line *SD* is calculated as

$$\theta = tan^{-1}\frac{(y_2 - y_1)}{(x_2 - x_1)} \qquad (2)$$

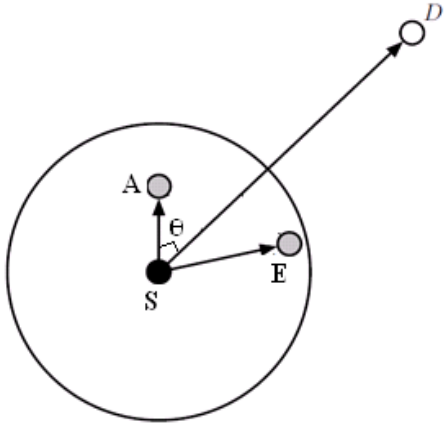

Fig. 3 Angle calculated by DIR Scheme

In Fig. 4, node *S* and *D* are source and destination nodes. Node *A* is a next-hop node of source node *S* because node *A* has minimum angle between itself, source node *S,* and the destination node *D*. Therefore, node *A* is selected as the next-hop forwarding node when it receives the message from *S*. It uses the same method, to find the next-hop forwarding node with minimum angle with line *SD*. In this way, node *B* is selected as a next-hop node of *A* for forwarding packets to destination. Finally node *C* directly delivers the packet to destination node *D*. During the whole D-LAR process route request packet is forwarded using LAR scheme.

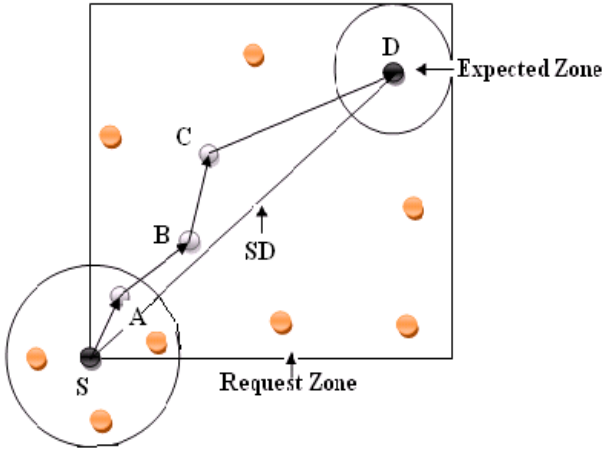

Fig. 4 Directional- location aided routing scheme

The D-LAR scheme is more suitable for dense vehicular network such as city traffic scenario, where an adequate number of vehicles in the city road at any given time in order to make connectivity between vehicles.

## 4. Feasibility Evaluation of VANET using D-LAR

The number of vehicles in the vehicular network has a crucial impact on network performance, when a high number of nodes are available in the network, which will increase the probability of finding the most suitable node to forward the packet to its desired destination. If there is link breakage has occurred, an alternative node can be selected immediately with a minimum time delay and the packet delivery ratio will increase when increasing the number of nodes in the network.

In VANET, vehicles can move into many different traffic environments that can interfere with wireless communication. Velocities of vehicles are restricted by speed limits, road traffic condition, and traffic control mechanism. Usually, there are large numbers of vehicles in city roads available at any given time. Take the metropolitan city Delhi in India for example, where plenty of vehicles like car, truck, buses, motorcycles etc. run on the road at any given time. Delhi has huge population and with respect to population, vehicles population in Delhi is large among all metropolitan cities in India. On an average about 800 new vehicles are added in Delhi every day [13]. According to Department of Transportation (DOT) annual reports, there are 50 thousand to 100 thousand of vehicles in the city in day times. Therefore, vehicle density can vary in day times. A number of road-based routing protocols have been designed to study the feasibility of VANET. In this work, the feasibility evaluation of VANET has explained using D-LAR routing protocol.

In D-LAR, the process of selecting the next-hop node in the network has an impact on the connectivity and reliability of the network, and selecting the suitable next-hop node will increase the feasibility of obtaining a high network performance. Let *S* be the forwarding node. Since as per the D-LAR scheme, the next-hop node must be present in the request zone for further transmission. Therefore, search for a next-hop node of *S* is restricted to the one fourth of the circular region of the source node *S*. In case of multi-hop VANET, each vehicle must be within the range of at least one other vehicle to maintain the link in the network. Let *N* be the number of nodes located in the circular area are follows Possion distribution with parameter *λ*, where *λ* is the node density per unit area. We assume that each node has the same transmission and receiving range of *R*, therefore, the average number of nodes within a circular area can be calculated as:

*Number of nodes (N) = λ × Area of the region*

$$= \lambda \pi R^2$$

Similarly if *n* nodes are distributed according to Possion process in one fourth area of the circular region (shaded area) of *S*, which is the part of request zone, then the number of nodes *n* in the shaded area as shown in the Fig. 5 is

$$n = \lambda \pi R^2 / 4 \quad (3)$$

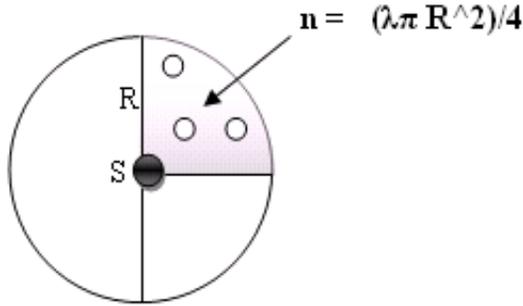

Fig. 5 Showing the shaded area of the circle

Since nodes are Possion distributed, then the probability that *n* other nodes located in the shaded area is given by [14]

$$P(n, A) = \frac{(\lambda A)^n \cdot e^{-\lambda A}}{n!}$$

$$= \frac{\left(\frac{\lambda \pi R^2}{4}\right)^n \cdot e^{-\frac{\lambda \pi R^2}{4}}}{n!}, \quad n = 0, 1, 2, 3, \ldots \quad (4)$$

Thus, the probability to select at least *k* out of *n* within the shaded area is

$$P_k = 1 - \sum_{n=0}^{k-1} \frac{\left(\frac{\lambda \pi R^2}{4}\right)^n \cdot e^{-\frac{\lambda \pi R^2}{4}}}{n!} \quad (5)$$

Figure 6 shows the probability of selecting at least *k* nodes in the transmission range of the source node. We have taken node density 0.0002 and 0.0004 nodes/km$^2$. In Fig. 7, the probability of selecting at least *k* nodes in the shaded area has shown for the same node densities. Based on these observations, we can say that the probability that a vehicle has at least *k* neighbors in the shaded area within the transmission range is relatively high.

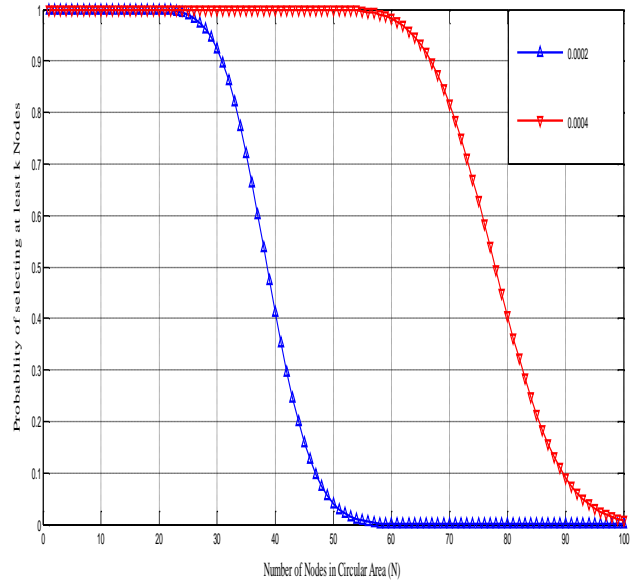

Fig. 6 Probability of selecting at least k nodes in the circular area

This study shows that shaded area belonging to request zone of the transmission range has less number of nodes compare to whole circular region of the source node. Less number of nodes in the shaded area improving the individual reachable neighbors will improve the connectivity between nodes in the network. Thus, Fig. 7 clearly shows that the probability to select a node as a next-hop node in shaded area is higher than the probability to select a node in whole circular area (as shown in Fig. 6) within the transmission range for further transmission.

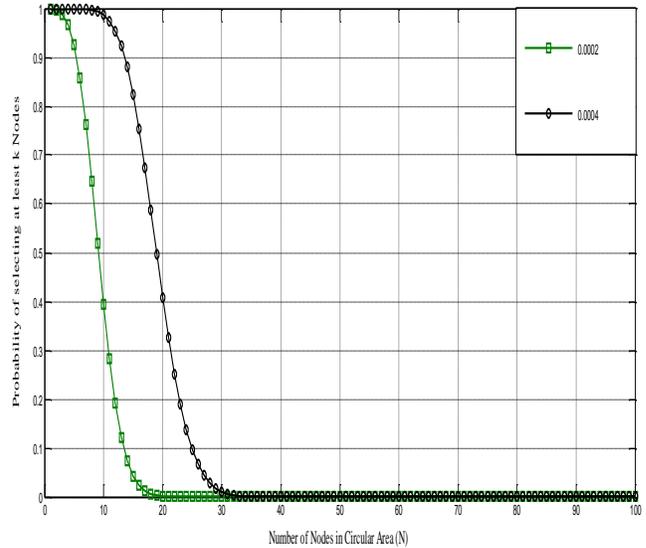

Fig. 7 Probability of selecting at least k nodes in the shaded area

## 5. Conclusion

In this paper, a new position-based routing protocol, D-LAR which uses the advantage of DIR and LAR routing protocols is proposed. In this scheme, the next-hop node is selected with minimum angle along with the straight line drawn between source and destination node in the requested zone as per DIR protocol. During the whole D-LAR packet transmission process route request packet is forwarded using LAR scheme. The D-LAR scheme is more suitable for dense vehicular network such as city traffic scenario, where an adequate number of vehicles in the city road at any given time in order to make connectivity between vehicles.

We studied the characteristics of VANET for dense network and the feasibility of VANET for D-LAR routing protocol have been justified. The performance of D-LAR protocol can be compared in terms of end-t-end delay, packet delivery ratio, and route costs with DIR and LAR routing protocols in the next step of work.

## References

[1] Hassnaa Moustafa and Yan Zhang, Vehicular networks: techniques, standards, and applications,Boca Raton London, New York:CRC Press, 2009.
[2] Jinhua Guo and Nathan Balon, "Vehicular Ad Hoc Networks and Dedicated Short-Range Communication," University of Michigan – Dearborn, 2006.
[3] Sun Xi, LI Xia-miao, "Study of the Feasibility of VANET and its Routing Protocols," IEEE, 2008, pp. 1-4.
[4] Kevin C. Lee, Uichin Lee, and Mario Gerla, " Survey of Routing Protocols in Vehicular Ad Hoc Networks," IGI Global, 2010, pp. 149-170.
[5] Hirozumi Yamaguchi, Weihua Sun, and Teruo Higashino, "Geographic Routing on Vehicular Ad Hoc Networks," IGI Global, 2010, pp. 171-178.
[6] Azzedine Boukerche, Horacio A.B.F. Oliveira, Eduardo F. Nakamura, and Antonio A.F. Loureiro, "Vehicular Ad Hoc Networks: A New Challenge for Localization-Based Systems," Computer Communication, Elsevier (ScienceDirect), 2008, pp.1-12.
[7] G. Dommety and R. Jain, "Potential networking applications of global positioning systems (GPS)," Tech. Rep. TR-24, CS Dept., The Ohio State University, April 1996.
[8] Young-BaeKo and Nitin H. Vaidya, "Location-aided routing (LAR) in mobile ad hoc networks," ACM/IEEE, MOBICOM'98, 1998, pp. 66-75.
[9] I. Stojmenovic, A. P. Ruhil, and D. K. Lobiyal, "Voronoi diagram and convex hull based Geocasting and routing in wireless networks", Wireless Communications and Mobile Computing Special Issue on Ad Hoc Wireless Networks, John Wiley & Sons, Ltd., Vol.6, Issue.2, 2006,pp.247-258.
[10] E. Kranakis, H. Singh and J. Urrutia, "Compass routing on geometric networks," Proceedings of the 11th Canadian Conference on Computational Geometry (CCCG'99), 1999, http:www.cs.ubc.ca/ conferences/CCCG/elec_proc/c46.ps.gz.
[11] David B. Johnson and David A. Maltz, "Dynamic Source Routing in Ad Hoc Wireless Networks," Kluwer Academic Publishers, 1996.
[12] Neng-Chung Wang, Jong-Shin Chen, Yung-Fa Huang, and Si-Ming Wang, "A Greedy Location-Aided Routing Protocol for Mobile Ad Hoc Networks," Proceedings of the 8th WSEAS International Conference on Applied Computer and Applied Computational Science, 2002, pp. 175-180.
[13] Rites Ltd. Urban environmental engineering household survey, Delhi, 1994.
[14] SUN Xi, LI Xia-miao, "Study of the Feasibility of VANET and its Routing Protocols", IEEE 4th International Conference on Wireless Communications, Networking and Mobile Computing (WOCOM), Dalian, 2008, pp.1-4.

**Dr. Ram Shringar Raw** received his B. E. (Computer Science and Engineering) from G. B. Pant Engineering College, Pauri-Garhwal, UK, India and M. Tech (Information Technology) from Sam Higginbottom Institute of Agriculture, Technology and Sciences, Allahabad (UP), India in 2000 and 2005, respectively. He has obtained his Ph.D (Computer Science and Technology) from School of Computer and Systems Sciences, Jawaharlal Nehru University, New Delhi, India in 2011. He is currently working as Assistant Professor in the Department of Computer Science and Engineering, Ambedkar Institute of Advanced Communication Technologies & Research, GGSIP University, New Delhi, India. His current research interest includes Mobile Ad hoc Networks and Vehicular Ad hoc Networks. Dr. Raw has published papers in International Journals and Conferences including IEEE, Springer, Inderscience, American Institute of Physics, AIRCC, etc.

**Mr. Sanjoy Das** did his B. E. and M.Tech in Computer Science. He has submitted his Ph.D in Computer Science, School of Computer and Systems Sciences, Jawaharlal Nehru University, New Delhi, India. He has worked as an Assistant Professor, Department of Computer Science and Engineering in G. B. Pant Engineering College, Uttarakhand , India. Also, he has worked as an Assistant Professor in the department of Information Technology, School of Technology, Assam University (A Central University), Silchar, Assam, India. Presently, working as an Assistant Professor, in Computer Science and Engineering Department, School of Engineering and Technology, Galgotias University, Greater Noida, UP, India. His current research interest includes Mobile Ad hoc Networks and Vehicular Ad hoc Networks, Distributed Systems.

**Dr. Nanhay Singh,** working as Associate Professor in Ambedkar Institute of Advanced Communication Technologies & Research, Govt. of NCT, Delhi-110031 (Affiliated to Guru Gobind Singh Indraprastha University, Delhi) in the Department of Computer Science & Engineering. He received his Ph.D (Computer Science and Technology) & M. Tech. (Computer Science & Engineering) from the Kurukshetra, University, Kurukshetra, Haryana. He has rich experience in teaching the classes of Graduate and Post-Graduate in India. He has contributed to numerous International journal & conference publications in various areas of Computer Science. He published more than 11 Research Paper in International Journals and Conferences. . He has also written an International book Titled as "Electrical Load Forecasting Using Artificial Neural Networks and Genetic Algorithm", in Global Research Publications New Delhi (India). His area of interest includes Distributed System, Parallel Computing, Information Theory & Coding, Cyber Law, and Computer Organization.

**Mr. Sanjeet Kumar received** his B. E. (Computer Science and Engineering) from IET, Dr. Bhimrao Ambedkar University, Agra, India in 2005. He is pursuing M. Tech. from Integral University Lucknow. He has five years of experience in teaching as well as in industry. His area of interest includes Mobile Computing, Ad hoc Network and Sensor Network.

**Mr. Shailender Kumar** obtained his B.E. (CSE) from MDU Rohtak and M. Tech. from Rajasthan University. He has more than 10 years of experience in teaching at various esteemed engineering colleges in India, like Delhi College of Engineering, Netaji Subhas Institute of Technology etc. Currently he is working as Assistant Professor (Pre-revised) at AIACTR, Delhi, India. His area of interest is Databases, Compiler Construction, Computer Network, Ad hoc Network etc.